\documentstyle[aps]{revtex}

\begin{document}
\author{Yong-Jun Liu$^{1,2}$ and Chang-De Gong$^{3,1}$}
\title{The study of spin-spin correlations in quasi-one-dimensional Heisenberg
antiferromagnetic clusters}
\address{$^1$National Key Laboratory of Solid States of Microstructure, Nanjing\\
University, Nanjing 210093, PRC\\
$^2$Complexity Science Center, Yangzhou University, Yangzhou 225002, PRC\\
$^3$CCAST (World Laboratory), P.O. Box 8730, Beijing 100080, China}
\maketitle

\begin{abstract}
For a class of quasi-one-dimensional clusters, by using exact
diagonalization, we study the effect of side spins on the spin-spin
correlations on chain. Our calculations show that the side spins added in
the same sublattice can effectively strengthen the spin-spin correlations in
large distance region and make the change tend to flat. It is exactly proved
that periodically adding side spins can set up magnetic long-range orders in
the ground state. Also we investigate the effect of the density of side
spins on correlation strength. The case that two sublattices have different
localized spins is discussed.
\end{abstract}

Spin-spin correlation and magnetic long-range order (LRO) are of fundamental
importance for quantum spin systems. They have been studied by many
analytical and numerical works. For the bipartite Heisenberg
antiferromagnetic (AF) systems, the spin-spin correlations of any two sites
in the ground state (GS) always are AF, {\it i.e.} the correlation functions
are positive when two sites belong to the same sublattice whereas negative
the different sublattices.\ However, it is not sufficient to set up the AF
LRO. The dimensionality of system is one of the key factors. It has been
rigorously proved that there exists N\'{e}el order in the GS \cite{Dyson} of
the three-dimensional system, whereas no N\'{e}el order for the
one-dimensional one. Although there is no magnetic LRO for one- and
two-dimensional (1D and 2D) Heisenberg antiferromagnets at temperature $T>0$
by Mermin-Wager theorem \cite{Mermin}, $T=0$ may be the critical point. When
the localized spin $S\geq 1$, it has been proved that the GS of the 2D
Heisenberg antiferromagnet has N\'{e}el order \cite{JN}. Notwithstanding the
rigorous proof of N\'{e}el order when $S=\frac 12$ has not been established
until now, many numerical and analytical works support its existence at $T=0$
\cite{Reger}.

The situation of quasi-one-dimensional (QOD) systems may be diverse due to
their various geometric structures. The discovery of a heavy-fermion
phenomenon in the $Ce$-doped Neodymium cuperate \cite{Brugger93} has led to
an increasing interest in the study of strongly correlated electrons coupled
antiferromagnetically to magnetic moments \cite
{W.Zhang97,W.Zhang96,Moukouri,W.Zhang95}. W. Zhang {\it et al. }have
investigated the case of a single magnetic impurity and found that the
spin-spin correlation function between the impurity spin and spins in the
chain extends over long range \cite{W.Zhang96}. Obviously, the magnetic LRO
can not be established through adding a single impurity spin to an 1D chain.
The situation may change when impurity spins are periodically added to the
chain in a special way. In this paper, by using exact diagonalization, we
investigate a class of QOD Heisenberg AF clusters, in which there are some
impurity spins sit beside an 1D finite chain (we call them as side spins),
and explore the effects of side spins on spin-spin correlations. Our
numerical results indicate that the spin-spin correlations in the region of
large distances can be enhanced by adding side spins in the same sublattice.
When periodically adding side spins, the decay of spin-spin correlations
with distances slows down and becomes obviously flat in the range of large
distances. For infinite 1D chains, it is analytically proved that adding
side spins can set up magnetic LROs. Also, we investigate the variation of
magnetic LROs with the density of side spins, and find that the decay of
ferromagnetic (F) LRO is faster than that of AF LRO.

We investigate the spin-$\frac 12$ Heisenberg AF system with interactions of
nearest neighbors, whose Hamiltonian reads

\begin{equation}
H=J\sum_{({\bf i},{\bf j})}\vec{S}({\bf i})\cdot \vec{S}({\bf j})\text{,} 
\eqnum{1}
\end{equation}
where $J>0$. $({\bf i},{\bf j})$ denotes the sum over pairs of nearest
neighbors. The spin-spin correlation can be written as $\Delta
(R_i-R_j)\equiv \left\langle G\left| \vec{S}({\bf i})\cdot \vec{S}({\bf j}%
)\right| G\right\rangle $. Here, $R_i$ and $R_j$ are the coordinates of
sites $i$ and $j$ respectively, and $\left| G\right\rangle $ represents the
GS. Let $\overline{i}$ denote the site by the $i$th site of the chain. At
first, we consider a finite chain of $23$ sites under free boundary
conditions and study the effect of a single side spin on the spin-spin
correlations. By exact diagonalization, we calculate $\Delta (R_i-R_j)$ for
site 2 ({\it i.e.} $j=2$ and $i=3,4,5,...$), and the numerical results are
shown in Fig. 2, here the side spin is placed at site $\bar{2}$, $\bar{3}$
and $\bar{4}$ respectively (see Fig 1.(a)). When the side spin is added on
site $\bar{3}$, $\Delta (R_i-R_j)$ becomes weaker than that of the pure 1D
chain. Namely, the side spin weakens the spin-spin correlations. But, when
the side spin is at site $\bar{2}$ (or $\bar{4}$), $\Delta (R_i-R_j)$
becomes stronger in the region of large distances. In this case, the side
spin strengthens the spin-spin correlations for most of $(R_i-R_j)$. When a
single side spin is placed at sites $\bar{5}$, $\bar{7}$, $\bar{9}$ and $%
\overline{11}$ respectively, our calculations show that the effect of the
side spin is similar to that of the side spin at site $\bar{3}$. And, when
the side spin is placed at site $\bar{6}$, $\bar{8}$, $\overline{10}$ and $%
\overline{12}$ respectively, it is similar to that of the side spin at site $%
\bar{2}$.

Under the periodic boundary conditions, the case of a single side spin has
been studied by Monte Carlo simulations and the spin correlation function is
found to extend over long range \cite{W.Zhang96}. For such a system, the GS
always takes the lowest possible total spin (LTS) no matter where the side
spin is attached. But the situation for a finite chain under free boundary
conditions becomes little complicated. The finite chain can be divided into
two `sublattices' as doing for infinite systems. Sites $1$, $3$, $5$, ..., $%
21$ and $23$ belong to sublattice $A$, and sites $2$, $4$, $6$, ..., $20$
and $22$ sublattice $B$ (Fig. 1(a)). Adding the single side spin at
different sites can leads to the GS taking different values. When the side
spin sits by the site of sublattices $B$, it belongs to sublattice $A$. By
Lieb-Mattis theorem \cite{Lieb62}, the GS has the global spin 1 and is
3-fold degenerate. In other words, the GS takes a higher total spin than its
lowest possible value. But, when the side spin sits by the site on
sublattices $A$, it belongs to sublattice $B$. The GS takes the LTS zero.
Our calculations show that the effect of a single side spin depends on where
the side spin is. In other words, the total spin of the GS is related to the
behaviors of spin-spin correlations. One important question concerned is how
the spin correlations change when more side spins are added. When the side
spins added are in the same sublattice, the total spin of the GS becomes
higher, and the difference between it and the lowest possible total spin
larger. From the above calculations, we speculate that $\Delta (R_i-R_j)$
will become stronger in the region of large distances.

For further details, we consider the cases of two side spins. If adding two
side spins at site $\bar{2}$ and $\bar{4}$, the GS has the global spin $%
\frac 32$ while the lowest possible total spin is $\frac 12$. But, if adding
two side spins by site $2$ and $3$, the GS takes the lowest possible total
spin $\frac 12$. From the above notion, one can speculate that spin-spin
correlations will become stronger if adding two side spin at sites $\bar{2}$
and $\bar{4}$, whereas those weaker at sites $\bar{2}$ and $\bar{3}$. We
calculate the spin-spin correlations between site $2$ and others. The
numerical results are shown in Fig. 2. They agree with the above speculation.

Also, we calculate spin-spin correlations for the following three cases (see
Fig. 1(a)): 1). 4 side spins sitting at sites $\bar{2}$, $\bar{8}$, $%
\overline{14}$ and $\overline{20}$ on a finite chain of 21 sites; 2). 5 side
spins sitting at sites $\bar{2}$, $\bar{6}$, $\overline{10}$, $\overline{14}$
and $\overline{18}$ on a finite chain of 19 sites; 3). 8 side spins sitting
at sites $\bar{2}$, $\bar{4}$, $\bar{6}$, $\bar{8}$, $\overline{10}$, $%
\overline{12}$, $\overline{14}$ and $\overline{16}$ on a finite chain of 17
sites. Their ground states have the global spins $S^T=\frac 52$, $3$ and $%
\frac 92$ respectively. The numerical results are plotted in Fig. 3. Our
calculations show that periodically adding side spins in the same sublattice
can obviously enhance the spin-spin correlations in the large distance
region. And, as the density of side spins increasing, spin-spin correlations
become stronger and decay more slowly. Approximately, we can fit the
spin-spin correlations in exponential way, {\it i.e. }$\Delta
(R_i-R_j)=C\exp (-\left| R_i-R_j\right| /\zeta )$, here $C$ is coefficient
and $\zeta $ the correlation length. We calculate $\zeta $ by $\Delta
(R_9-R_2)$ and $\Delta (R_{15}-R_2)$. For the bare chain of 21 sites, $\zeta
=9.30$. But, for the above three cases, $\zeta =51.38$, $77.27$ and $7345.38$%
. It shows that the large distance behavior of spin-spin correlations is
enhanced by adding side spins. Especially, $\zeta \sim 10^3$ for the third
case. It is much larger than the correlation length of bare chain, even the
size of system. Consequently, side spins can slow down effectively the decay
and make the variation become flat in the region of large distances (Fig.
3). It seems to exist the AF LRO.

Now, we turn to consider an infinite QOD system, which is constructed by
periodically adding side spins in the same sublattice (Fig. 1(b)). The total
number of sites is $N=K(l+2)$, here $K$ denotes the number of cells and $l$
the number of sites on chain between every two side spins. $l$ must take
odd, {\it i.e.} $l=2k+1$, here $k=0,1,2...$. Supposing the side spin is in
sublattice $A$, the number of sites of sublattice $A$ is $N_A=K(l+3)/2$ and
that of sublattice $B$ $N_B=K(l+1)/2$. By the Lieb-Mattis theorem \cite
{Lieb62}, the global spin of the GS is

\begin{equation}
\Lambda =K\left| (k+2)S_A-(k+1)S_B\right| \text{,}  \eqnum{2}
\end{equation}
where $S_A$ and $S_B$ are the values of localized spins on sublattice $A$
and $B$ respectively. To investigate the existence of magnetic LRO, one
needs to calculate the quantity

\begin{equation}
g({\bf q})\equiv \left\langle G\left| \vec{S}(-{\bf q})\cdot \vec{S}({\bf q}%
)\right| G\right\rangle \text{.}  \eqnum{3}
\end{equation}
$\vec{S}({\bf q})=\frac 1{\sqrt{N}}\sum_{{\bf j}}\vec{S}({\bf j})\exp (i{\bf %
q\cdot j)}$, ${\bf q}$ is a reciprocal vector. The criterion of magnetic
LROs is that $g({\bf q})\geq O(N)$ at some ${\bf q}$. If $g({\bf Q})\geq
O(N) $, here ${\bf Q=(}\pi ,\pi ,...,\pi )$, there exists AF LRO. If $g({\bf %
0})\geq O(N)$, there exists F LRO. Following the approaches developed by G.
S. Tian \cite{Tian94,Tian97}, one can obtain

\begin{equation}
g({\bf Q)>}g({\bf 0)=}\frac{\Lambda ^2+\Lambda }N.  \eqnum{4}
\end{equation}
We discuss the case that the sites on two sublattices have the equal
localized spin, {\it i.e.} $S_A=S_A=S$. From equation (2), we readily obtain 
$\Lambda =NS/(l+2)$. As long as $l$ is finite, it is always true that

\begin{equation}
g({\bf Q)>}g({\bf 0)>}O(N).  \eqnum{5}
\end{equation}
From these inequalities, one can conclude that AF and F LROs coexist in the
GS, and the former is predominant. In other words, side spins set up
magnetic LROs.

Obviously, although there always exist magnetic LROs for finite $l$, both of
F and AF correlation strengths will depend on the density of side spins,
which is defined as $\eta \equiv 1/(l+2)$. We introduce the following two
quantities to measure F and AF correlation strengths respectively,

\begin{equation}
\Gamma _F\equiv \frac 1{N^2}\sum_{{\bf i,j}}\left\langle G\left| \vec{S}(%
{\bf i})\cdot \vec{S}({\bf j})\right| G\right\rangle  \eqnum{6}
\end{equation}
and

\begin{equation}
\Gamma _{AF}\equiv \frac 1{N^2}\sum_{{\bf i,j}}\lambda _{{\bf ij}%
}\left\langle G\left| \vec{S}({\bf i})\cdot \vec{S}({\bf j})\right|
G\right\rangle ,  \eqnum{7}
\end{equation}
where $\lambda _{{\bf ij}}=1$ when sites ${\bf i}$ and ${\bf j}$ belong to
the same sublattice and $\lambda _{{\bf ij}}=-1$ when sites ${\bf i}$ and $%
{\bf j}$ the different sublattices. $\Gamma _F$ and $\Gamma _{AF}$ will
decrease as $l$ increasing. When $l$ reaches to the infinite, the system
changes into an 1D chain and the magnetic LROs vanish ({\it i.e.} $\Gamma
_F=\Gamma _{AF}=0$). Possibly, the variation speeds of $\Gamma _F$ and $%
\Gamma _{AF}$ are different. We calculate the ratio $\rho \equiv \Gamma
_F/\Gamma _{AF}$ for $l=1$, $3$ and $5$, and plot the data in Fig. 4. $\rho $
decreases as $\eta $ decreasing approximately in the linear way. Then the
decay speed of F correlation strength is faster than that of AF correlation
strength. In other words, the ferrimagnetism becomes weaker and weaker as
the density of side spins decreasing.

Ref. \cite{Tian97} has proved that the GS of 1D Heisenberg AF chain has
magnetic LROs when its two sublattices have unequal localized spins. It is
interesting to investigate the QOD system with unequal localized spins. From
equation (2) and (4), we can conclude that if

\begin{equation}
\frac{S_A}{S_B}=\frac{k+1}{k+2}\text{,}  \eqnum{8}
\end{equation}
$g({\bf 0})=0$ due to $\Lambda =0$. It means that there is no F LRO. The
simplest case is $S_A=\frac 12$ , $S_B=1$ and $l=1$. One can give a spin
picture of the GS in valence-bond version. The spin on sublattice $B$ can be
divided into two $\frac 12$ spins. One of them forms a singlet with the
nearest side spin, and the other combines with its nearest neighbor on
chain. We think this kind of configurations governs the physics of the GS.
And it is responsible for the F LRO to vanish. Although we have not exactly
proved that the AF LRO can not exist in the GS of such systems, we believe
it is true. But, if $S_A/S_B\neq (k+1)/(k+2)$, the AF and F LROs coexist in
the GS.

We would like to thanks Prof. G. S. Tian for suggestions. The part of
calculations in this work have been done on the SGI Origin 2000 in the Group
of Computational Condensed Matter Physics, National Laboratory of Solid
State Microstructures, Nanjing University. This work was partially supported
by the Ministry of Science and Technology of China under Grant No.
NKBRSF-G19990646.

\smallskip {\LARGE Figure Captions:}

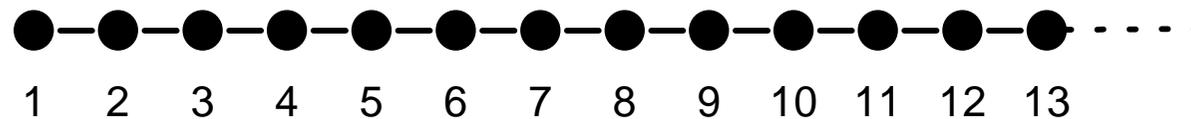
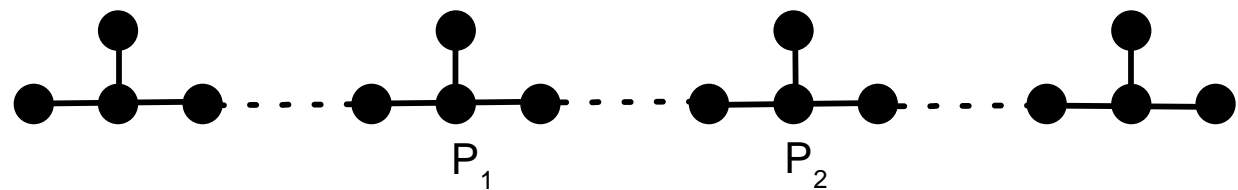
{\bf Fig. 1: }(a) the 1D Heisenberg AF chain. (b) the QOD Heisenberg AF
chain. There are $l$ sites between site $P_1$ and $P_2$, and $l=2k+1$, here $%
k=0,1,2...$.

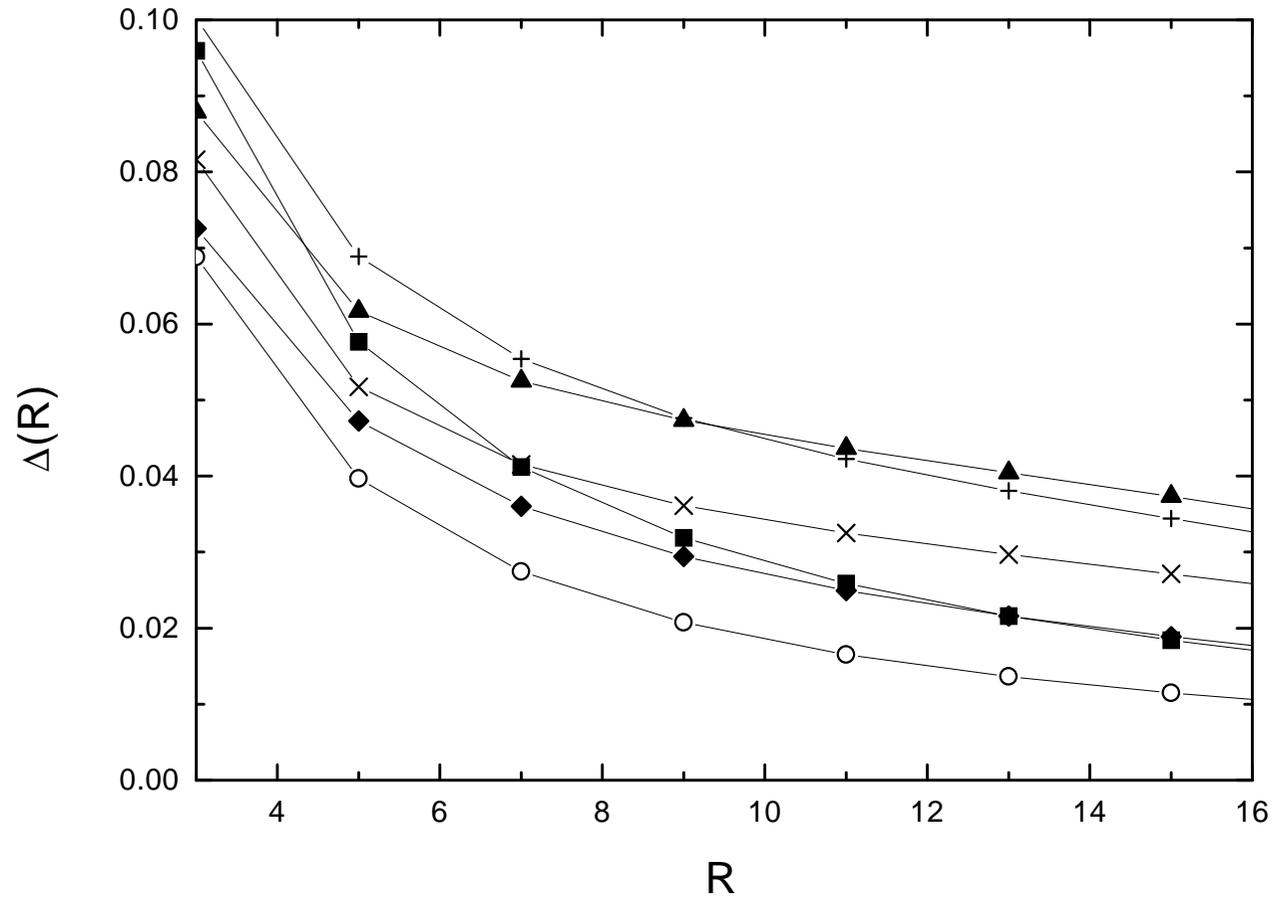
{\bf Fig. 2: }Spin-spin correlations between site 2 and others (see Fig.
1(a)) for these cases: no side spin (square); single side spin at site $%
\overline{2}$ (cross $+$), $\overline{3}$ (circle) and $\overline{4}$ (cross 
$\times $) respectively; two side spins at sites $\overline{2}$ and $%
\overline{4}$ (triangle); two side spins at sites $\overline{2}$ and $%
\overline{3}$ (diamond). Here, $R=R_i-R_2$. The length of chain is 23.

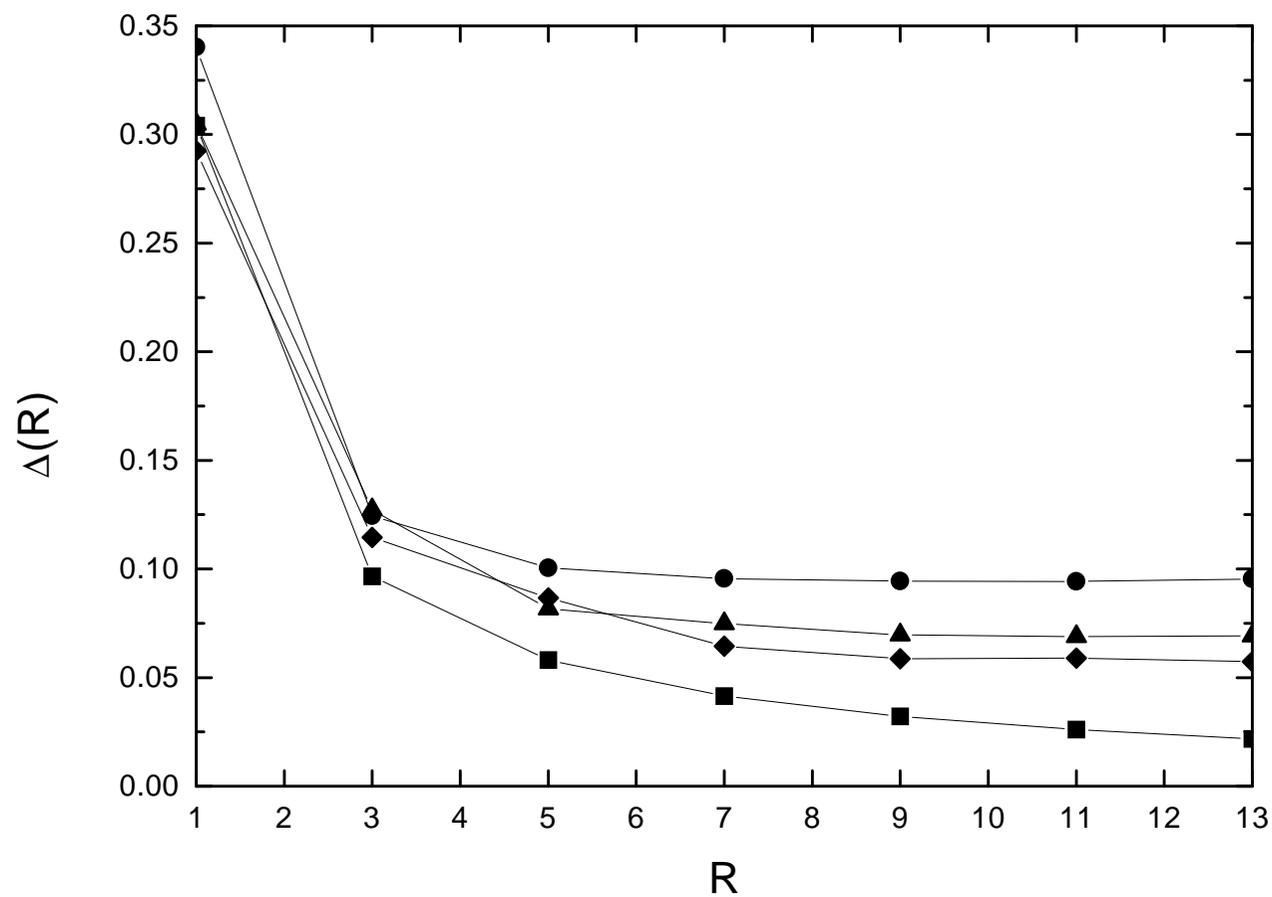
{\bf Fig. 3: }Spin-spin correlations between site 2 and others (see Fig.
1(a)) for these cases: bare chain of 21 sites (square); 4 side spins at
sites $\overline{2}$, $\overline{8}$, $\overline{14}$ and $\overline{20}$ on
a chain of 21 sites (diamond); 5 side spins by site $\overline{2}$, $%
\overline{6}$, $\overline{10}$, $\overline{14}$ and $\overline{18}$ on a
chain of 19 sites (triangle); 8 side spins by site $\overline{2}$, $%
\overline{4}$, $\overline{6}$, $\overline{8}$, $\overline{10}$, $\overline{12%
}$, $\overline{14}$ and $\overline{16}$ on a chain of 17 sites (solid
circle). Here, $R=R_i-R_2$.

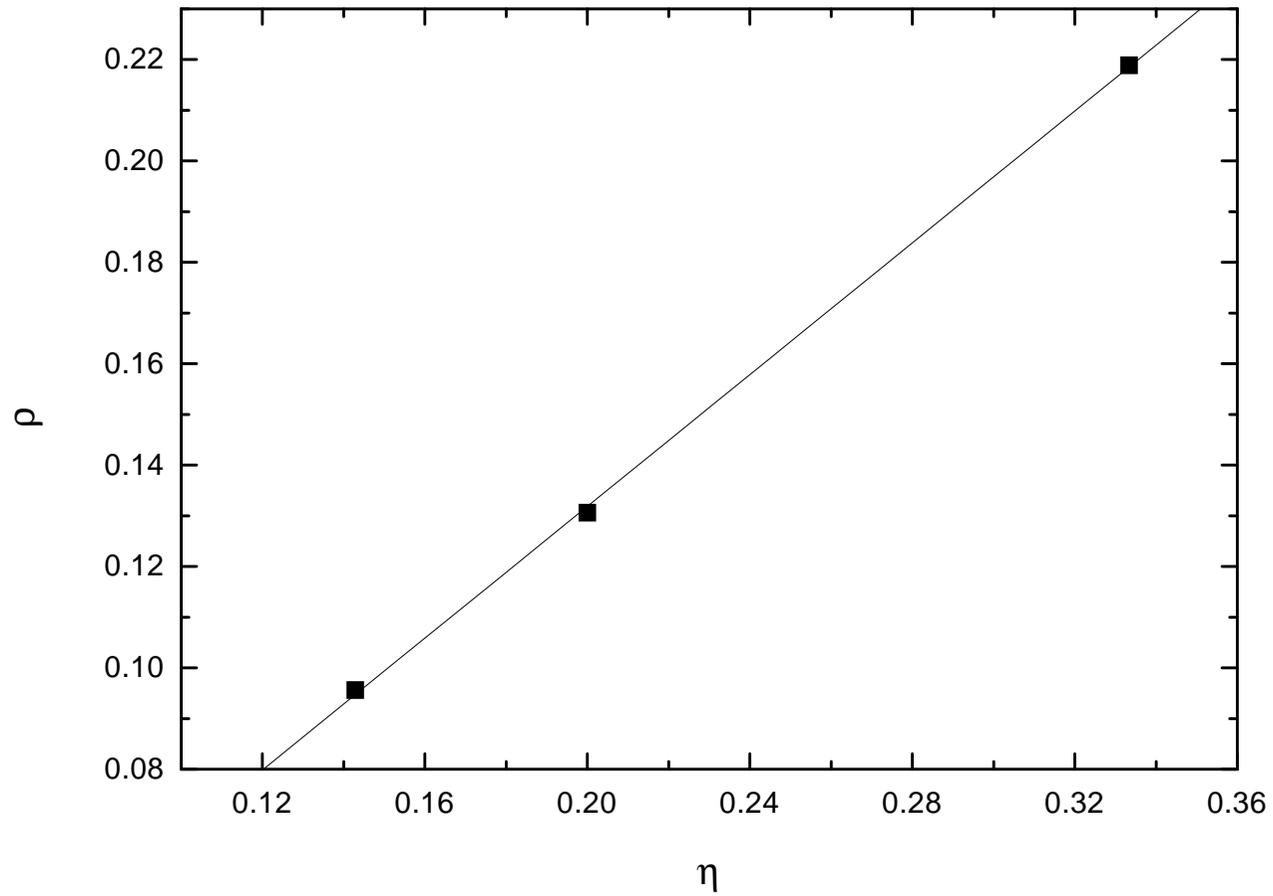
{\bf Fig. 4: }$\rho $ vs. $\eta $. The solid line is obtained by fitting $%
\rho $ in linear way.

\end{document}